\def \aap #1 #2 {{ Astron. Astrophys.\/} {\bf #1}, #2}
\def \aal #1 #2 {{ Astron. Astrophys. Lett.\/} {\bf #1}, L#2}
\def \aar #1 #2 {{ Astron. Astrophys. Rev.\/} {\bf #1}, #2}
\def \aas #1 #2 {{ Astron. Astrophys. Suppl. Ser.\/} {\bf #1}, #2}
\def \aj #1 #2 {{ Astron. J.\/} {\bf #1}, #2}
\def \annrev #1 #2 {{ Ann. Rev. Astron. Astrophys.\/} {\bf #1}, #2}
\def \apj #1 #2 {{ Astrophys. J.\/} {\bf #1}, #2}
\def \apjl #1 #2 {{ Astrophys. J. Lett.\/} {\bf #1}, L#2}
\def \apjs #1 #2 {{ Astrophys. J. Suppl.\/} {\bf #1}, #2}
\def \apss #1 #2 {{ Astrophys. Space Sci.\/} {\bf #1}, #2}
\def \ASR #1 #2 {{ Adv. Space Res.\/} {\bf #1}, #2}
\def \BAIC #1 #2 {{ Bull. Astron. Inst. Czechosl.\/} {\bf #1}, #2}
\def \JSQRT #1 #2 {{ J. Quant. Spectrosc. Radiat. Transfer\/} {\bf #1}, #2}
\def \mnras #1 #2 {{ Mon. Not. R. astr. Soc.\/} {\bf #1}, #2}
\def \MEM #1 #2 {{ Mem. R. astr. Soc.\/} {\bf #1}, #2}
\def \PLR #1 #2 {{ Phys. Rev. Lett.\/} {\bf #1}, #2}
\def \pasj #1 #2 {{ Publ. Astron. Soc. Japan\/} {\bf #1}, #2}
\def \PASP #1 #2 {{ Publ. Astr. Soc. Pacific\/} {\bf #1}, #2}
\def \nat #1 #2 {{ Nature\/} {\bf #1}, #2}
\def \SAIT #1 #2 {{ Mem.\ Soc.\ Astron.\ It.\/} {\bf #1}, #2}
\def \MESS #1 #2 {{ The Messenger\/} {\bf #1}, #2}
\def \ASTRNACH #1 #2 {{ Astron. Nach.\/} {\bf #1}, #2}

\documentclass[12pt,a4paper]{article}
\usepackage[top=2cm, bottom=2cm, left=2cm, right=2cm]{geometry}
\usepackage{graphicx}% Include figure files
\usepackage{bm}% bold math
\usepackage{amsmath,amssymb}
\newcommand{\eqb}{\begin{equation}}
\newcommand{\eqe}{\end{equation}}
\begin{document}

%\preprint{AIP/123-QED}

\title{Asymptotic theory of relativistic, magnetized jets}

\author{Yuri Lyubarsky\\
Physics Department, Ben-Gurion University}
 \date{}

%\pacs{52.30.Cv, 98.54.Aj, 98.70.Rz}% PACS, the Physics and Astronomy
                             % Classification Scheme.
%\keywords{relativistic magnetohydrodynamics, quasars, gamma-ray bursts}%Use showkeys class option if keyword %display desired
\maketitle
\centerline{\bf Abstract}
The structure of a relativistically hot, strongly magnetized jet is investigated at large distances from the source. Asymptotic equations are derived describing collimation and acceleration of the externally confined jet. Conditions are found for the transformation of the thermal energy into the fluid kinetic energy or into the Poynting flux. Simple scalings are presented for the jet collimation angle and Lorentz factors.

\vskip 1cm
\section{Introduction}

Highly collimated, relativistic outflows are commonly observed in compact
astronomical objects (gamma-ray bursts, active galactic nuclei,
galactic black-hole and neutron-star binaries). It is widely believed that in all these cases, the jet is driven by rotating, twisted magnetic fields
%mechanism is related to magnetic  fields
(see, e.g., recent reviews in ref.\cite{jets}). The rapidly spinning central body (neutron star, accretion disk, black hole) twists up the magnetic field into a toroidal component and the plasma is ejected by the magnetic tension. In relativistic flows, the energy per particle significantly exceeds the rest mass energy therefore in order to create a relativistic jet, one has to convey a significant energy to a small amount of the matter. The main  advantage of the magnetic launch mechanism is that the magnetic field lines, like driving belts, could in principle transfer the rotational energy to a low density periphery of the central engine thus forming a baryon pure but energetic outflow. The relativistic velocities could potentially be achieved if the magnetic energy density in the plasma frame exceeds the plasma energy density. In such outflows, the energy is transported, at least initially, in the form of the Poynting flux. The question is how and where the electromagnetic energy is eventually transformed to the plasma energy.

In the scope of ideal MHD, the energy could be transferred to the plasma only via gradual acceleration by electromagnetic stresses. In non-relativistic flows, the plasma is accelerated centrifugally when sliding
along the rotating poloidal field lines. The azimuthal field is generated only when the plasma inertia becomes comparable with the magnetic stresses so that the field lines have to bend backwards. This implies that a good fraction of the Poynting flux is converted into the kinetic energy of the flow already when the azimuthal field becomes comparable with the poloidal one. In relativistic, highly magnetized flows, the magnetic force is generally balanced not by inertia but by the electric force. The azimuthal field becomes comparable with the poloidal one, both of them being comparable with the electric field, at the light cylinder defined as a surface on which the corotational velocity is equal to the speed of light (in differentially rotating magnetospheres, this surface is not a cylinder but
we retain the standard term, which has come from the pulsar theory). The fluid kinetic energy remains small at the light cylinder. Beyond the light cylinder, the conservation of the magnetic flux implies that the poloidal magnetic field decreases as $1/r^2$, where $r$ is the cylindrical radius of the jet. The azimuthal field and the electric field decrease only as $1/r$ being close to each other so that the fluid is only slowly accelerated by a small residual force. By this reason, the acceleration zone is extended well beyond the light cylinder so that formation of relativistic jets spans a very large range of scales. In recent studies, both numerical and analytical \cite{komissarov07,komissarov09,tchekhovskoy08,tchekhovskoy09,tchekhovskoy10,lyubarsky09,lyubarsky10}, the general conditions were formulated for collimation and acceleration of relativistic MHD jets and the efficiency of the Poynting flux into the kinetic energy conversion was thoroughly examined.   Non-steady Poynting dominated outflows have also being studied \cite{granot_etal10,levinson10,lyutikov10a,lyutikov10b}. In these works, only cold flows have been addressed. Here we relax this assumption and study relativistically hot, highly magnetized jets. By relativistically hot we mean the fluid with the pressure exceeding the rest mass energy density, $p>\rho c^2$. Such outflows are believed to be formed in gamma-ray bursts (see, e.g., reviews \cite{piran04,meszaros06}). Here we study such outflows in the far zone where most of acceleration occurs.

When the flow expands, the fluid cools down. In the pure hydrodynamical case, the thermal energy is converted into the kinetic energy of the fluid. It follows immediately from the Bernoulli equation that the relativistically hot fluid  with the adiabatic index $\Gamma=4/3$ is accelerated such that the flow Lorentz factor grows proportionally to the jet radius. In a magnetized flow, the thermal energy could be converted not only into the kinetic energy but also into the Poynting flux. The aim of this study is to explore the fate of the thermal energy in expanding, Poynting dominated jets. We consider outflows confined by the pressure of the external medium because only in this case the jet could be collimated. The paper is organized as follows. In section 2, the general equations governing ideal relativistic axisymmetric flows are presented. In section 3, asymptotic theory is developed describing such outflows in the far zone. In sect. 4, the theory is applied to narrow jets from rigidly rotating sources. Conclusions are presented in sect. 5.

\section{Relativistic MHD equations for a steady axisymmetric flow}

For the sake of consistency and in order to introduce
notations, let us shortly review the basic theory of relativistic,
magnetized outflows \cite{okamoto78,lovelace_etal86,li_chiueh_begelman92}.
In order to describe ideal MHD flows one has to employ the Maxwell equations, the flux-freezing condition and the continuity, entropy and momentum equations. For relativistic, steady state flows these equations are written as
 \eqb
\nabla\cdot\mathbf{B}=0;\quad \mathbf{\nabla\times E}=0;
 \label{maxwell1}\eqe
 \eqb
 \mathbf{\nabla\times B}=4\pi\mathbf{j};\quad \mathbf{\nabla\cdot E}=4\pi\rho_e;
 \eqe
  \eqb
 \mathbf{E}+\mathbf{v\times B}=0.
 \label{freezing}\eqe
 \eqb
\nabla\cdot(\rho\gamma\mathbf{v})=0; \quad \mathbf{v\cdot\nabla}\left(\frac p{\rho^{\Gamma}}\right)=0;
 \label{cont}\eqe
 \eqb
\rho\gamma(\mathbf{v\cdot\nabla})(h\gamma\mathbf{v})=-\nabla p+
\rho_e\mathbf{E}+\mathbf{j\times B}.
 \label{motion}\eqe
Here  $\mathbf{v}$ is the velocity of the flow and $\gamma\equiv(1-v^2)^{-1/2}$ the associated Lorentz factor,
$\mathbf{E}$ and $\mathbf{B}$ the electric and magnetic fields as measured in the lab frame,
$\rho_e$ and $\mathbf{j}$ the charge and current densities also in the lab frame, and
$\rho$, $h$ and $p$ the plasma proper mass density, specific enthalpy and the pressure. The speed of light is taken to be unity throughout the paper. The above equations should be complemented by the equation of state, which we assume to be polytropic with
the index $\Gamma$. Then the specific enthalpy may be presented as
 \eqb
h=1+\frac{\Gamma}{\Gamma-1}\frac p{\rho}.
 \label{enthalpy}\eqe
Here the first term represents the rest mass energy.

In axisymmetrical flows, the magnetic field is
conveniently decomposed into the poloidal and azimuthal
components, %which will be denoted by subscripts $p$ and $\phi$, correspondingly.
$\mathbf{B}_p$ and $B_{\phi}$.
The poloidal field is expressed via the flux function as
 \eqb
 \mathbf{B}_p=\frac 1r\nabla\Psi\times\mathbf{\widehat{\phi}}.
 \label{Bfield}\eqe
We use cylindrical $(r,\phi,z)$ coordinates; the hat denotes
unit vectors. The flux freezing condition (\ref{freezing}) implies that the magnetic surfaces are equipotentials
so that one can write
  \eqb
E=r\Omega(\Psi) B_p;
 \label{EBp}\eqe
where $\Omega$ is the angular velocity of the magnetic field line. The position of the light cylinder is found from the condition $\Omega r=1$.
The plasma streams along the magnetic surfaces so that one can decompose the fluid velocity into the poloidal (along $\mathbf{B}_p$) and azimuthal components. Then Eqs. (\ref{freezing}) and (\ref{EBp}) yield
 \eqb
B_pv_{\phi}-B_{\phi}v_p=r\Omega(\Psi) B_p;
 \label{Omega}\eqe
which implies that the fluid slides along the magnetic field lines rigidly rotating with the angular velocity $\Omega$.
It follows immediately from Eqs. (\ref{EBp}) and (\ref{Omega}) that well beyond the light cylinder, $\Omega r\gg 1$,
 the electric and azimuthal magnetic field are close to each other and both are much larger than the poloidal field, $E\approx B_{\phi}\approx\Omega rB_p$.

If the shape of the flux surfaces is given, the flow is completely described by five integrals of motion. The first one is the angular velocity of the field line, $\Omega(\Psi))$, defined by Eq. (\ref{Omega}). Three others are the mass, energy and angular momentum-to-magnetic flux ratios,
 \eqb
\eta(\Psi)=\frac{4\pi\rho v_p\gamma}{B_p};
 \label{continuity}\eqe
 \eqb
\mu(\Psi)= h\gamma-\frac{r\Omega B_{\phi}}{\eta}
 \label{energy}\eqe
 \eqb
l(\Psi)= h\gamma rv_{\phi}-\frac{rB_{\phi}}{\eta}.
 \label{momentum}\eqe
The fifth integral is the specific entropy
\eqb
q(\Psi)=\frac p{\rho^{\Gamma}}.
\label{entropy}\eqe
These equations are obtained by projecting Eqs. (\ref{cont}) and (\ref{motion}) onto the direction of the flow, $\mathbf{\widehat{l}}=\mathbf{\widehat{n}\times\widehat{\phi}}$, where
$\mathbf{\widehat{n}}=\nabla\Psi/\vert\nabla\Psi\vert$ is the normal to the flux surfaces.
The integrals of motion are generally found by prescribing the conditions at the inlet of the flow and from the condition of smooth passage of the solution through the critical points of the equations \cite{li_chiueh_begelman92,tsinganos_etal96,bogovalov97,
vlahakis_etal00}. Here we consider the flow in the far zone, outside of all critical points, therefore all integrals of motion  are considered in this paper as given functions of $\Psi$.

Making use of the above conservation laws, one can express any physical quantity via, e.g.,
$\gamma$, $h$ and $\Psi$. For example,
  \eqb
 v_{\phi}=\frac 1{\Omega r}\left(1-\frac{\mu-\Omega l}{h\gamma}\right).
 \label{vphi}\eqe
If the rotation velocity  at the origin of the outflow is well below the speed of light, $\Omega r_{\rm in},v_{\phi,{\rm in}}\ll 1$, one can write Eq. (\ref{vphi}) in the form
 \eqb
v_{\phi}=\frac 1{\Omega r}\left(1-\frac{h_{\rm in}\gamma_{\rm
in}}{h\gamma}\right);
 \label{v_phi}\eqe
where the index "in" is referred to the parameters of the injected
plasma. Below we will use this formula because it results in a bit simpler expressions than the general
formula (\ref{vphi}). In the general case, we should just substitute $h_{\rm in}\gamma_{\rm
in}$ by $\mu-\Omega l$ in all the following expressions. One can also found straightforwardly $v_p$ as a function of $\gamma$, $h$ and $\Psi$. Then the identity $v_p^2+v_{\phi}^2+\gamma^{-2}=1$ may be written as the Bernoulli equation
 \begin{eqnarray}
&&\frac{\Omega^4r^4B_p^2}{\eta^2(\mu-h\gamma)^2}
\left[1-\frac 1{\Omega^2 r^2}\left(1-\frac{h_{\rm in}\gamma_{\rm
in}}{h\gamma}\right)\right]^2 \nonumber\\
&&+\frac 1{\Omega^2 r^2}\left(1-\frac{h_{\rm in}\gamma_{\rm in}}{h\gamma}\right)^2+\frac
1{\gamma^2}=1;
 \label{Bernoulli}\end{eqnarray}
which connects the Lorentz factor of the flow with the geometry of the flux tube defined by
the flux function $\Psi$.

In order to get a complete set of equations, one has to make projection of the equation of motion onto the normal to the flux surface. This yields the transfield
force-balance equation (the generalized Grad-Shafranov equation)
 \begin{eqnarray}
&&\frac 1{\cal
R}\left[h\rho\gamma^2v_p^2+\frac{E^2-B_p^2}{4\pi}\right]+
\frac 1{r^2}\left(h\rho\gamma^2v_{\phi}^2+\frac{B_p^2}{4\pi}\right)\mathbf{\widehat n}\cdot\mathbf{r}
\nonumber \\
&&-\mathbf{\widehat n}\cdot\nabla p=\frac1{8\pi r^2}\mathbf{\widehat n}\cdot\nabla
\left[r^2(B^2-E^2)\right].
 \label{transfield}\end{eqnarray}
Here $\cal R$ is the local curvature radius of the poloidal
field line (defined such that $\cal R$ is positive when the flux surface is concave so that the collimation
angle decreases)
 \eqb
\frac 1{\cal R}\equiv -\mathbf{\widehat{n}\cdot}(\mathbf{\widehat{l}\cdot\nabla})\mathbf{\widehat{l}}=
\mathbf{\widehat{n}\cdot}[\mathbf{\widehat{l}\times}(\mathbf{(\mathbf{\nabla\times\widehat{l}})})]=
-\mathbf{\widehat{\phi}\cdot}(\mathbf{\nabla\times\widehat{l}}).
 \eqe
The transfield equation (\ref{transfield}) is reduced to an equation for $\gamma$, $h$ and $\Psi$ with the aid of Eqs. (\ref{enthalpy}) and (\ref{EBp}) -- %, (\ref{Omega}), (\ref{continuity}), (\ref{energy}), (\ref{momentum}) and
(\ref{entropy}). %Therefore Eqs. (\ref{Bernoulli}) and (\ref{transfield}) form a complete set of equations.

\section{Relativistic jet in the far zone}

In this section, we derive asymptotic equations describing the flow at large distances from the source.
For cold relativistic flows the procedure was developed in ref. \cite{lyubarsky09}; here we present the straightforward generalization to the hot case. We are going to obtain the Bernoulli and transfield equations in the limit
$\Omega r\gg 1$, $\gamma\gg 1$. These two conditions are not independent for outflows that are initially Poynting dominated because such outflows are generally accelerated therefore such outflows are in any case highly relativistic in the far zone.

The zeroth approximation in $(\Omega r)^{-1}$ and $\gamma^{-1}$ to the Bernoulli equation is obtained by taking the limit $\Omega r$, $\gamma\to\infty$ in Eq. (\ref{Bernoulli}), which yields
 \eqb
\eta(\mu-h\gamma)=\Omega^2r^2B_p.
 \label{Bernoulli0}\eqe
This equation could be written, with account of Eqs. (\ref{EBp}) and
(\ref{energy}), as
 \eqb B_{\phi}+E=0.
 \label{zeroth_order}\eqe
In the transfield equation (\ref{transfield}), the leading
order in $1/r$ and $1/\gamma$ terms are those in the right-hand side, $\mathbf{n}\cdot\nabla(B^2_{\phi}-E^2)$, because the terms
in the left-hand side are small either as $B_p/B_{\phi}\sim (\Omega r)^{-1}$ or as
$r/\cal R$ (the curvature of the flow lines in the far zone is small). This means that if one  substitutes Eq. (\ref{zeroth_order}) into the right-hand side of the
transfield equation, one would kill the leading order terms.
The correct procedure \cite{vlahakis04,lyubarsky09} is to expand the
Bernoulli equation (\ref{Bernoulli}) to the first non-vanishing
order in $1/r$ and $1/\gamma$ and only then to eliminate the
leading order terms from Eq.(\ref{transfield}).

Expanding Eq.
(\ref{Bernoulli}) yields
 \eqb
B^2-E^2%\equiv\left(\frac{\eta(\mu-\gamma)}{\Omega r}\right)^2-(\Omega r)^2B_p^2
=\frac{\Omega^2r^2B^2_p}{\gamma^2}\left[1+\left(\frac{h_{\rm in}\gamma_{\rm in}}{h\Omega r}
\right)^2\right].
 \label{Bernoulli1}\eqe
Substituting this relation into the right-hand side of Eq.
(\ref{transfield}), one comes to
%  \begin{eqnarray}
%\frac 1{\cal
%R}\left[h\rho\gamma^2v_p^2+\frac{E^2-B_p^2}{4\pi}\right]+
%\frac 1{r^2}\left(h\rho\gamma^2v_{\phi}^2+\frac{B_p^2}{4\pi}\right)\mathbf{\widehat n}\cdot\mathbf{r}
%-\mathbf{\widehat n}\cdot\nabla p \nonumber \\ =
%\frac1{8\pi r^2}\mathbf{\widehat n}\cdot\nabla\left\{
%\frac{\Omega^2r^4B^2_p}{\gamma^2}\left[1+\left(\frac{h_{\rm in}\gamma_{\rm in}}{h\Omega r}
%\right)^2\right]\right\}.
% \label{tr} \end{eqnarray}
an equation that does not contain terms which nearly cancel each other. Therefore one can
safely simplify this equation further out. Namely, we can now  neglect $B_p$ as compared with $E$, substitute unity instead of $v_p$ and also use the Bernoulli equation in the form (\ref{Bernoulli0}) in order to transform terms.
The procedure is the same as that was described in sect. 4.1 of ref. \cite{lyubarsky09}. The resulting asymptotic transfield equation is written as
  \begin{eqnarray}
&&\frac{\eta\mu B_p}{\cal R}+\frac{\eta B_p}{\Omega^2 r^4}
\left(\mu-2h_{\rm in}\gamma_{\rm in}+
\frac{h_{\rm in}^2\gamma_{\rm in}^2}{\gamma h}\right)\mathbf{\widehat n}\cdot\mathbf{r}
-\mathbf{\widehat n}\cdot\nabla p
\nonumber \\
&&=\frac 1{2r^2}\mathbf{\widehat n}\cdot\nabla
\left\{\frac{\eta^2(\mu-h\gamma)^2}{\Omega^2\gamma^2}\left[1+\left(\frac{h_{\rm in}\gamma_{\rm in}}{h\Omega r}
\right)^2\right]\right\}.
 \label{tr1} \end{eqnarray}
% \eqb
%r^2\left(B^2-E^2\right)=\frac{\eta^2(\mu-h\gamma)^2}{\Omega^2\gamma^2}\left[1+\left(\frac{h_{\rm in}\gamma_{\rm in}}{h\Omega r}
%\right)^2\right]
%\label{tr2} \eqe

 In the most interesting case of collimated flows, $z\gg r$, one
can take $\mathbf{\widehat n}\cdot\mathbf{r}=r$ and
$\mathbf{\widehat n}\cdot\nabla=\partial/\partial r$. When
looking for the shape of the magnetic surfaces, one can
conveniently use the unknown function $r(\Psi,z)$ instead of
$\Psi(r,z)$. Then, e.g.,
 \eqb
B_p=\frac 1r\vert\nabla\Psi\vert\approx\frac 1r\frac{\partial\Psi}{\partial r}=
\left(r\frac{\partial r}{\partial\Psi}\right)^{-1}.
 \label{Bp_jet}\eqe
In the same approximation, the curvature radius may be presented as
(note that $\cal R$ is defined to be positive for concave surfaces)
 \eqb
\frac 1{\cal R}=-\frac{\partial^2r}{\partial z^2}.
 \label{curvature}\eqe
Now the transfield equation for the collimated flows in the far
zone could be written as
  \begin{eqnarray}
&&\eta\mu\left[-\frac{\partial^2 r}{\partial z^2}+\frac 1{\Omega^2 r^3}
\left(1-\frac{2h_{\rm in}\gamma_{\rm in}}{\mu}+
\frac{h_{\rm in}^2\gamma_{\rm in}^2}{\mu\gamma h}\right)\right]-r\frac{\partial p}{\partial\Psi}
\nonumber \\
&&=\frac 1{2r}\frac{\partial}{\partial\Psi}
\left\{\frac{\eta^2(\mu-h\gamma)^2}{\Omega^2\gamma^2}\left[1+\left(\frac{h_{\rm in}\gamma_{\rm in}}{h\Omega r}
\right)^2\right]\right\}.
 \label{coll_transfield} \end{eqnarray}
%where the expression in the right-hand side is given by Eq. (\ref{tr2}).
This equation should be supplemented by the zeroth order Bernoulli equation (\ref{Bernoulli0}),
which is written in the new variables as
\eqb
(\mu-h\gamma)\frac{\partial r}{\partial\Psi}=\frac{\Omega^2}{\eta}r.
\label{Bernoulli2}\eqe

In order to get a closed set of equations, one has to
express the pressure and the enthalpy via $r$ and $\gamma$. In the far zone, $v_p\to 1$, $v_{\phi}\ll 1$, the continuity equation (\ref{cont}), together with
Eqs. (\ref{Omega}) and (\ref{energy}), gives
 \eqb
\rho=\frac{\eta B_p}{4\pi\gamma}=\left(\frac{\eta}{\Omega r}\right)^2\frac{\mu-h\gamma}{4\pi\gamma}.
 \eqe
Now one can write the entropy equation (\ref{entropy}) as
 \eqb
p=p_{\rm in}\left(\frac{r^2_{\rm in}\gamma_{\rm in}}{r^2\gamma}\frac{\mu-h\gamma}{\mu-h_{\rm in}\gamma_{\rm in}}\right)^{\Gamma}
 \label{pressure}\eqe
whereas the equation of state (\ref{enthalpy}) is written in the form
 \eqb
\frac{h-1}{h_{\rm in}-1}=\left(\frac{r^2_{\rm in}\gamma_{\rm in}}{r^2\gamma}\frac{\mu-h\gamma}{\mu-h_{\rm in}\gamma_{\rm in}}\right)^{\Gamma-1}.
 \label{enthalpy1}\eqe

Equations (\ref{coll_transfield}) and (\ref{Bernoulli2}), supplemented by Eqs. (\ref{pressure}) and (\ref{enthalpy1}), form a complete set of equations describing relativistic, hot, magnetized, collimated outflows in the far zone. We believe that these equations suit well to numerical solution
because they do not contain terms that nearly cancel each
other. However they also could be used in order to find simple analytical scalings in all limiting cases.

Here we are interested in outflows subtending a finite magnetic flux $\Psi_0$.
% therefore Eqs. (\ref{coll_transfield}) and (\ref{Bernoulli0}) should be solved at $0\le\Psi\le\Psi_0$.
If the flow is confined by the pressure of the external medium, $p_{\rm ext}(z)$, the pressure
balance condition should be satisfied at the boundary:
 \eqb
\frac 1{8\pi}(B^2-E^2)+p=p_{\rm ext}(z).
 \eqe
Making use of Eqs. (\ref{Bernoulli0}) and (\ref{Bernoulli1}), one writes the boundary condition
in the form
 \eqb
\left\{\frac{\eta^2(\mu-h\gamma)^2}{8\pi r^2\Omega^2\gamma^2}\left[1+\left(\frac{h_{\rm in}\gamma_{\rm in}}{h\Omega r}
\right)^2\right]+p\right\}_{\Psi=\Psi_0}=p_{\rm ext}(z).
 \label{boundary} \eqe
%The boundary condition at the axis of the flow is just $r(\Psi=0)=0$.

\section{Relativistically hot Poynting dominated jet}
In this section, we consider in detail a relativistically hot flow,
 \eqb
 h\gg 1.
 \label{cond_h}\eqe
which was launched as Poynting dominated,
\eqb
h_{\rm in}\gamma_{\rm in}\ll\mu.
\label{cond_Poynting}\eqe
When the flow expands, the plasma cools so that eventually the condition (\ref{cond_h}) is violated.
The condition (\ref{cond_Poynting}) ensures that the flow still remains Poynting dominated at this point
therefore the condition
 \eqb
h\gamma\ll\mu
 \label{condition}\eqe
is fulfilled everywhere in the domain of interest.
The enthalpy equation (\ref{enthalpy1})  is now reduced to
 \eqb
\frac{h}{h_{\rm in}}=\left(\frac{r^2_{\rm in}\gamma_{\rm in}}{r^2\gamma}\right)^{1/3}.
 \label{enthalpy2} \eqe
From here on we will use the relativistic adiabatic index $\Gamma=4/3$. The expanding accelerating fluid cools down so that that the condition (\ref{cond_h}) could be written as
\eqb
 r^2\gamma\ll h_{\rm in}^3r_{\rm in}^2\gamma_{\rm in}.
 \label{cool-radius}\eqe

For simplicity, we consider the jet with a constant angular velocity, $\Omega=\it const$ and homogeneous injection, $\eta=\it const$; $\gamma_{\rm in}=\it const$. In this case, one can conveniently use the dimensionless
variables
 \eqb
X=\Omega r;\quad Z=\Omega z.
 \eqe
We also assume
that the energy integral, $\mu(\Psi)$, is described by a linear function
  \eqb
\mu(\Psi)=h_{\rm in}\gamma_{\rm
in}\left(1+\frac{\Psi}{\widetilde{\Psi}}\right);\quad
\widetilde{\Psi}=\frac{h_{\rm in}\gamma_{\rm in}\eta}{2\Omega^2}.
 \label{linear_psi}\eqe
The last represents in fact the expansion of $\mu(\Psi)$ to the first order in $\Psi$ so that this expression is universally correct close to the axis of the flow (see sect. 3 in ref. \cite{lyubarsky09}). Numerical simulations show that in jets with a constant angular velocity, the linear function (\ref{linear_psi}) is a good approximation to $\mu$ all the way to the boundary of the flow $\Psi=\Psi_0$  (see, e.g., fig. 4 in ref.\cite{komissarov07}). In this expression, the first and the second terms describe the kinetic and
the Poynting energy flux,
correspondingly. Note that the Poynting flux goes to zero at the axis therefore the flow is Poynting dominated only at $\Psi\gg\widetilde{\Psi}$; in this domain, the energy integral is reduced to
\eqb
\mu(\Psi)=\frac{2\Omega^2\Psi}{\eta}.
\label{energy1}\eqe

\subsection{The governing equation}
We are going to solve Eqs. (\ref{coll_transfield}) and (\ref{Bernoulli2}) under the conditions (\ref{cond_h}) and (\ref{condition}). The condition (\ref{condition}) implies that one can neglect the second term in curly brackets in the left-hand side of the transfield equation (\ref{coll_transfield}). The third term in the curly brackets is small at the inlet of the flow however, the product $h\gamma$
could in principle grow with the distance. By virtue of Eq. (\ref{enthalpy2}) this is possible only if
$\gamma$ grows faster than $r$. We will see that this never happens therefore we neglect this term.

Let us also demonstrate that the last term in the left-hand side of Eq. (\ref{coll_transfield}) is small as compared with the right-hand side of this equation. Making use of Eq. (\ref{Bernoulli0}) one can estimate the ratio of these terms as
$p\gamma^2/(\Omega^2r^2B_p^2)$. The expression in the numerator is of the order of the plasma energy flux
whereas the expression in the denominator is, by virtue of Eqs. (\ref{EBp}) and (\ref{zeroth_order}), of the order
of the Poynting flux. Therefore the above ratio is small.

Now the transfield equation (\ref{coll_transfield}) is written, with the aid of Eq. (\ref{energy1}), as
  \eqb
\Psi X\left(-\frac{\partial^2 X}{\partial Z^2}+\frac 1{X^3}
%\left[1+\frac{h_{\rm in}\gamma_{\rm in}}{\mu}\left(\frac{\gamma_{\rm in}r}{\gamma r_{\rm in}}\right)^{2/3}\right]
\right)=
\frac{\partial}{\partial\Psi}\left\{\frac{\Psi^2}{\gamma^2}
\left[1+\left(\frac{\gamma\gamma_{\rm in}^{2}}{XX_{\rm in}^{2}}\right)^{2/3}\right]\right\}.
\label{Poynting_jet}\eqe
%\eqb
%X^2\left(B^2-E^2\right)=\frac{4\Omega^4\Psi^2}{\gamma^2}
%\left[1+\left(\frac{\gamma\gamma_{\rm in}^{2}}{XX_{\rm in}^{2}}\right)^{2/3}\right].
% \label{B-E}\eqe
In the same approximation, the boundary condition (\ref{boundary}) is reduced to
 \eqb
\left\{\frac{\Omega^4\Psi^2}{X^2\gamma^2}
\left[1+\left(\frac{\gamma\gamma_{\rm in}^{2}}{XX_{\rm in}^{2}}\right)^{2/3}\right]\right\}_{\Psi=\Psi_0}=8\pi p_{\rm ext}(z).
 \label{boundary1} \eqe

At the condition (\ref{condition}) and with account of Eq. (\ref{energy1}) one reduces the Bernoulli equation (\ref{Bernoulli2}) to an equation for $X(\Psi,Z)$:
 \eqb
2\Psi\frac{\partial X}{\partial\Psi}=X.
 \eqe
The general solution to this equation may be presented as
 \eqb
X=R(Z)\sqrt{\frac{\Psi}{\Psi_0}}.
\label{x(psi)}\eqe
%\eqb
%\Phi(\Psi)=\sqrt{2}\exp\left(\int_{\widetilde{\Psi}}^{\Psi}\frac{\Omega^2d\Psi}{\mu\eta}\right);
% \eqe
where $R(z)$ is an arbitrary function, which is in fact the radius of the jet.
%Substituting Eq. (\ref{energy1}), one gets
% \eqb
% \Phi(\Psi)=\sqrt{1+\frac{\Psi}{\widetilde{\Psi}}}\approx\sqrt{\frac{\Psi}{\widetilde{\Psi}}};
% \label{Phi}\eqe
One sees that the structure of collimated, Poynting dominated
jets is generally self-similar. Recall that this
equation is valid only at $\Psi\gg\widetilde{\Psi}$.
%The quantity $D$ is roughly the radius of the very inner
%part of the jet, $\Psi\sim\widetilde{\Psi}$.
Note that Eq. (\ref{x(psi)}) implies that the poloidal magnetic field is homogeneous,
$\Psi\propto X^2$; $\partial B_p/\partial X=0$; such a distribution is indeed seen
in numerical simulations\cite{tchekhovskoy08}.

In order to find the function $R(z)$, let us substitute
Eq.(\ref{x(psi)}) into the left-hand side of Eq. (\ref{Poynting_jet}) and integrate
the obtained equation between $\widetilde{\Psi}$ and $\Psi_0$:
  \begin{eqnarray}
&&\Psi_0^2\left(-\frac{1}{3}R\frac{d^2R}{dZ^2}+\frac 1{R^2}\right) \\
&&=\left\{\frac{\Psi^2}{\gamma^2}
\left[1+\left(\frac{\gamma\gamma_{\rm in}^{2}}{XX_{\rm in}^{2}}\right)^{2/3}\right]\right\}^{\Psi=\Psi_0}_{\Psi=\widetilde{\Psi}}.
  \nonumber\end{eqnarray}
One can take the expression in the right hand side only at the upper limit because we will see that $\gamma$ grows not faster than $X\propto\sqrt{\Psi}$ so that this expression grows with $\Psi$. Making use of the boundary condition (\ref{boundary1}) one gets a closed equation for the jet shape:
 \eqb
-\frac{1}{3}R\frac{d^2R}{dZ^2}+\frac 1{R^2}=\frac{8\pi p_{\rm ext}R^2}{\Psi_0^2\Omega^4}.
 \eqe
 %One can conveniently introduce a new variable
% \eqb
% Y=3^{-1/4}\left(\frac{\Psi_0}{\widetilde{\Psi}}\right)^{1/2}D.
% \label{Y}\eqe
%It follows immediately from Eqs. (\ref{x(psi)}) and (\ref{Phi})
%that $Y$ is equal, to within the factor $3^{1/4}$, to the external radius of the jet.
Presenting the external pressure as
 \eqb
p_{\rm ext}(Z)=p_0{\cal P}(Z);
 \eqe
where $p_0=p_{\rm ext}(1)$ is the external pressure extrapolated to the light cylinder one presents this equation in a dimensionless form
  \eqb
\frac{d^2R}{dZ^2}-\frac 3{R^3}+\beta {\cal P}(Z)R=0.
 \label{Y(Z)}\eqe
We introduced the parameter
\eqb
\beta=\frac{6\pi p_0}{\Omega^4\Psi^2_0}=\frac{6\pi p_0}{B_0^2};
 \label{beta_lin}\eqe
where
$B_0\equiv \Omega^2\Psi_0$ is the characteristic magnetic field
at the light cylinder.

Having found the jet shape, one can easily
find the full structure of the jet therefore Eq. (\ref{Y(Z)}) may be called the governing equation.
If $R(Z)$ is found from the governing equation, the shape of all the flux surfaces is immediately obtained from Eq. (\ref{x(psi)}).
%, (\ref{Phi}) and (\ref{Y}) as
 %\eqb
%X(Z,\Psi)=3^{1/4}\sqrt{\frac{\Psi}{\Psi_0}}Y(Z).
% \label{X}\eqe
In order to find the Lorentz factor of the
flow, one can substitute Eq. (\ref{x(psi)}) into Eq. (\ref{Poynting_jet})
and then perform integration from 0 to $\Psi$, which yields an equation for $\gamma$
 \eqb
 -\frac{\Psi}{3\Psi_0}R\frac{d^2R}{dZ^2}+\frac{\Psi_0}{\Psi}\frac 1{R^2}=
\frac{1}{\gamma^2}\left[1+\frac{\Psi_0}{\Psi}
\left(\frac{\gamma\gamma_{\rm in}^2}{RR^2_{\rm in}}\right)^{2/3}\right]
 \label{Lorentz-factor}\eqe
%One can easily find $\gamma$ from this equation provided $Y(Z)$ is known.

It is interesting that the governing equation (\ref{Y(Z)}) coincides, to within change of variables, with that for cold flows \cite{komissarov09,lyubarsky09}.
Therefore the results for cold jets could be straightforwardly generalized to the hot case.
If the external pressure is distributed according to a power law,
 \eqb
{\cal P}=\frac 1{Z^{\kappa}},
 \label{P}\eqe
the general solution to the governing equation (\ref{Y(Z)}) could be found analytically (see ref. \cite{lyubarsky09}).
Below we describe these solutions. Since behavior of the solutions at large $Z$
depends on the sign of $\kappa-2$, we consider different cases separately.

\subsection{The case $\kappa<2$.}
If $\kappa<2$, the flow expands as\cite{tchekhovskoy08,komissarov09,lyubarsky09}
 \eqb
R=\left(\frac{3Z^{\kappa}}{\beta}\right)^{1/4}.
 \label{Ykappa<2}\eqe
This asymptotics could be found directly from Eq. (\ref{Y(Z)}) by
neglecting the first term. Substituting Eq. (\ref{Ykappa<2}) into Eq.(\ref{Y(Z)}), one sees that  at $\kappa<2$, the first term is really less than the second one.
According to Eq. (\ref{x(psi)}), all flux surfaces are similar to the surface of the jet boundary
therefore in the transfield equation (\ref{coll_transfield}), one can also neglect the term $\partial^2 r/\partial z^2$. Then the transfield equation becomes an ordinary differential equation describing cylindrical equilibria. This means that the structure of the jet at any $z$ is the same as the structure of an appropriate cylindrical equilibrium configuration, parameters of the configuration being adiabatically changed with $z$.
Such a collimation regime is designated in ref. \cite{lyubarsky09} as the equilibrium regime.
%Neglect of the term $d^2R/dZ^2$ in the governing equation
%so that the collimation occurs in the equilibrium regime.

The general
solution at $\kappa<2$ also expands as $Z^{\kappa/4}$
but very long wave oscillations are superimposed on this
expansion \cite{lyubarsky09}. These are in fact free oscillations around the
equilibrium state arising if the flow was injected not in the equilibrium state.
They do not affect the stability of the flow. The amplitude of
these oscillations could be found by matching to the near zone
solution at $Z\sim 1$. The spatial period of these oscillations
increases with the distance as $Z^{\kappa/2}$.

The Lorentz factor of the flow is found from Eq.
(\ref{Lorentz-factor}). For the smooth expansion described by
Eq. (\ref{Ykappa<2}) one can neglect the first term in the
left-hand side, which yields
  \eqb
\frac 1{\gamma^2}\left[1+\frac{\Psi_0}{\Psi}
\left(\frac{\gamma\gamma_{\rm in}^2}{RR^2_{\rm in}}\right)^{2/3}\right]=\frac{\Psi_0}{\Psi}\frac 1{R^2}.
 \eqe
The solution to this equation could be presented as
 \eqb
\gamma=a(\Psi)R
\eqe
where the function $a(\Psi)$ satisfies the equation
   \eqb
1+\frac{\Psi_0}{\Psi}\left(\frac{a\gamma_{\rm in}^2}{R^2_{\rm in}}\right)^{2/3}=\frac{\Psi_0}{\Psi}a^2.
 \label{eq_a}\eqe
%In cold Poynting dominated flows, the corresponding asymptotics is $\gamma=R$.
In relativistically hot, nonmagnetized flows, the Lorentz factor grows proportionally to the flow radius
\eqb
\gamma=\gamma_{\rm in}R/R_{\rm in};
 \label{gamma}\eqe
which is followed directly from the Bernoulli equation.
The last expression is also valid well inside the light cylinder of the Poynting dominated flow because at $R\ll 1$, the magnetic field is nearly force-free and the rotation velocity is small so that one can neglect the centrifugal acceleration. In this case the plasma just flows along the expanding field lines and the expression (\ref{gamma}) ensures that the flow becomes highly relativistic, $\gamma\gg 1$, already at $R=1$ provided the radius of the central body is much less than the light cylinder radius. This implies $\gamma_{\rm in}\gg R_{\rm in}$; then Eq. (\ref{eq_a}) yields $a=\gamma_{\rm in}/R_{\rm in}$. Now one sees that the hydrodynamical expression (\ref{gamma}) is valid also for the magnetized flow in case $\kappa<2$.
% Taking into account that $a>1$, one sees that
%the Lorentz factor of the relativistically hot flow  grows faster than in the case of the cold flow
%but still follows the general law $\gamma\propto X$, which is valid also in hot, non-magnetized flows.
This in fact means that the thermal energy is transformed into the kinetic energy of the flow.

It follows from Eqs. (\ref{cool-radius}), (\ref{Ykappa<2}) and (\ref{gamma}) that the fluid could be considered as relativistically hot only at distances less than
 \eqb
Z_c=\left[\frac{\beta}3h_{\rm in}^4\left(\frac{R_{\rm in}\gamma_{\rm in}}a\right)^{4/3}\right]^{1/\kappa}.
 \eqe
Beyond this distance the shape of the flow is still described by Eq. (\ref{Ykappa<2}) whereas the Lorentz factor grows as $\gamma=R$ until the flow ceases to be Poynting dominated \cite{lyubarsky09}.

\subsection{The case $\kappa=2$.}
If $\kappa=2$, the solution to Eq. (\ref{Y(Z)}) is written as \cite{komissarov09,lyubarsky09}
 \eqb
R=\left(\frac{3}{\beta-1/4}\right)^{1/4}Z^{1/2};\qquad \beta>1/4;
 \label{Ykappa=2a}\eqe
  \eqb
R=CZ^{(1+\sqrt{1-4\beta})/2};\qquad \beta<1/4.
 \label{Ykappa=2b}\eqe
Let us consider these two cases separately.

The $\beta>1/4$ solution is similar to the equilibrium solution (\ref{Ykappa<2}) and goes to this solution as $\beta\gg 1/4$. Note that the solution (\ref{Ykappa<2}) goes to $Y\propto Z^{1/2}$ as $\kappa\to 2$ therefore the solution (\ref{Ykappa=2a}) could be considered as a limiting case $\kappa\to 2$.

Substituting the solution (\ref{Ykappa=2a}) to Eq. (\ref{Lorentz-factor}), one gets the equation for the Lorentz factor of the flow. Just as in the previous section, the solution could be presented as
 \eqb
\gamma=a\sqrt{Z};\qquad \beta>1/4;
 \eqe
where $a$ satisfies the equation
 \eqb
1+\frac{\Psi_0}{\Psi}\left[\left(\frac{\beta-1/4}3\right)^{1/4}\frac{a\gamma_{\rm in}^2}{R^2_{\rm in}}\right]^{2/3}=
\frac{\Psi_0}{\Psi}\frac{\beta a^2}{\sqrt{3(\beta-1/4)}}.
 \eqe
The condition $\gamma_{\rm in}\gg R_{\rm in}$ (see the previous section) implies
 \eqb
\gamma=\frac{3^{1/4}\sqrt{\beta-1/4}}{\beta^{3/4}}\frac{\gamma_{\rm in}}{R_{\rm in}}\sqrt{Z}.
 \eqe
So in this case the acceleration regime is close to that for the equilibrium flow (\ref{gamma}) extrapolated
to $\kappa=2$.

Now let us consider the $\beta<1/4$ solution (\ref{Ykappa=2b}). The constant $C$ in this solution is not defined; it could be found only by matching to the near zone solution. If the flow was not
collimated at $Z\sim 1$, there should be $C\sim 1$.
The $\beta<1/4$ flow is collimated slower than $Y\propto Z^{1/2}$; this solution could be obtained by neglecting the second term in the left-hand side of the governing equation (\ref{Y(Z)}) and, correspondingly,
in the tramsfield equation (\ref{coll_transfield}). Such a flow was called in ref.\cite{lyubarsky09} non-equilibrium
because in this case, the pressure of the poloidal magnetic field is negligible small so that the
flow could be conceived as composed from coaxial magnetic loops.

In order to find the Lorentz factor in the case $\beta<1/4$ let us substitute the solution (\ref{Ykappa=2b}) to Eq. (\ref{Lorentz-factor}). One sees that one can neglect the second term in the left-hand side. Inspection of the obtained equation shows that $\gamma$ should grow slower than $R$ so that one can eventually neglect the second term
in the square brackets. Then one finally finds
 \eqb
\gamma=\frac 1C\sqrt{\frac{3\Psi_0}{\beta\Psi}}\,Z^{(1-\sqrt{1-4\beta})/2};\qquad \beta<1/4.
 \label{g1}\eqe
This asymptotics coincides with that for the cold flow \cite{komissarov09,lyubarsky09}, which in fact means that the thermal energy is not transformed into the kinetic energy. In the relativistically hot flow,
the specific plasma energy flux could be presented, with the aid of Eq. (\ref{enthalpy2}), as $h\gamma\propto (\gamma/r)^{2/3}$. In the non-magnetized plasma, $h\gamma=\it const$ therefore $\gamma\propto r$. In the case under consideration, $h\gamma$ decreases with the distance, which means, according to the energy equation
(\ref{energy}) that the plasma thermal energy is converted into the Poynting flux.

According to Eqs. (\ref{cool-radius}), (\ref{Ykappa=2b}) and (\ref{g1}), the flow becomes cool at the distance
 \eqb
Z_c=\left[\sqrt{\frac{\beta}3}\frac{h_{\rm in}^2\gamma_{\rm in}}C\right]^{2/(3+\sqrt{1-4\beta})}.
 \eqe
Beyond this point, the flow shape and the Lorentz factor are still described by Eqs. (\ref{Ykappa=2b}) and (\ref{g1}), correspondingly until the flow ceases to be Poynting dominated \cite{lyubarsky09}.

\begin{figure*}
\includegraphics[scale=0.5]{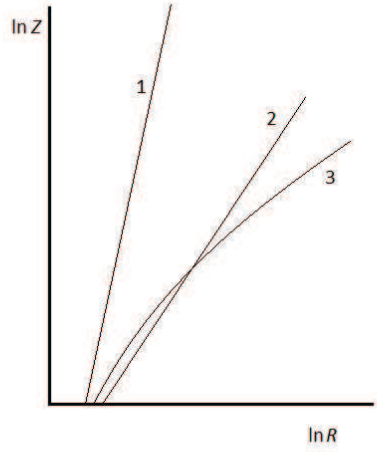}
\includegraphics[scale=0.5]{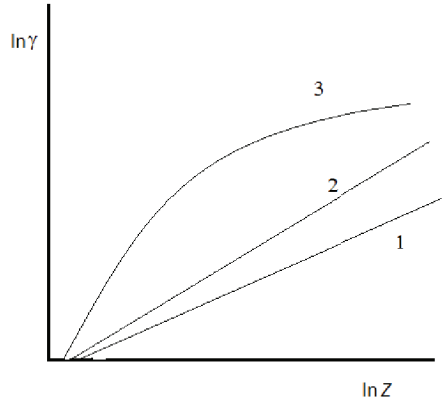}
\caption{Sketches of the jet radius (a) and Lorentz factor (b) as functions of the distance from the origin. The curve 1 describes the case $\kappa<2$; both the jet radius and the Lorentz factor grow as a power law with the index less than 1/2, see Eqs. (\ref{Ykappa<2}) and (\ref{gamma}) respectively. The case $\kappa=2$, $\beta>1/4$ may be considered as the limiting case of this regime with both the radius and the Lorentz factor growing $\propto \sqrt{Z}$.
The case $\kappa=2$, $\beta<1/4$ is shown by curves 2; the jet expands according to Eq. (\ref{Ykappa=2b}) faster than $\sqrt{Z}$ whereas the Lorentz factor grows according to Eq. (\ref{g1}) slower than $\sqrt{Z}$. The curve 3 is for $\kappa$ slightly larger than 3; at the initial stage (in the region (\ref{Z0})) the jet is rapidly expanded (according to Eq. (\ref{Ykappa<2}))
and accelerated (according to Eq. (\ref{gamma})). Then it becomes conical preserving the acquired collimation angle (in logarithmic coordinates, this looks as if the jet inflates) whereas the acceleration rate decreases and is described by Eq. (\ref{gamma_kappa>2}).}
\end{figure*}

\subsection{The case $\kappa>2$.}
At $\kappa>2$, the solution to the governing equation (\ref{Y(Z)}) goes, at large $Z$, to a linear function
\eqb
R=\Theta Z;
 \label{cone}\eqe
which means that the flow becomes radial at large distances. The flow could be collimated, $R\ll Z$, if $\kappa$ only slightly exceeds 2 or/and $\beta$ is large;
the final collimation angle could be presented as\cite{lyubarsky09} (note that our definition of $\Theta$ differs from that in ref. \cite{lyubarsky09} by a factor $3^{1/4}$)
  \eqb \Theta=\frac
{3^{1/4}}{\sqrt{\pi}}\Gamma\left(\frac{\kappa-1}{\kappa-2}\right)
\left(\frac{(\kappa-2)^{\kappa}}{\beta}\right)^{1/[2(\kappa-2)]};
 \eqe
where $\Gamma(x)$ is the gamma function.
One sees that the collimation angle rapidly increases with
increasing $\kappa$ and decreasing $\beta$:
$\Theta=0.013/\beta^{2.5}$ at $\kappa=2.2$, $\Theta=0.26/\beta$
at $\kappa=2.5$ and $\Theta=0.74/\sqrt{\beta}$ at $\kappa=3$.
Thus the flow could be collimated only at $\kappa< 3$.

One can get a simpler expression for the collimation angle assuming that $\kappa-2$ is a small number and making use of Stirling's approximation for the gamma function; the resulting expression reads as
 \eqb
\Theta=\frac{12^{1/4}}{\left(e\sqrt{\beta}\right)^{1/(\kappa-2)}}.
 \label{Theta}\eqe
One can check that even at $\kappa=3$ this expression provides a 10\% accuracy.

It has been shown in ref. \cite{lyubarsky09} that the collimation occurs only in the region
 \eqb
Z<Z_0=\left[\frac{2\sqrt{\beta}}{\kappa-2}\right]^{2/(\kappa-2)}.
 \label{Z0}\eqe
In this region, the shape of the jet is described by Eq. (\ref{Ykappa<2}) whereas the Lorentz factor grows according to Eq. (\ref{gamma}). Therefore the $\kappa<2$ scalings remain valid at $\kappa>2$ but only till some limiting distance.
At $Z>Z_0$, the jet shape approaches the cone (\ref{cone}). The curvature of the flux surfaces decreases in this region; it could be estimated from the governing equation (\ref{Y(Z)}) as
 \eqb
\frac{d^2R}{d^2Z}=-\frac{\beta\Theta}{Z^{\kappa-1}}.
 \label{curv_vs_r}\eqe
Then Eq. (\ref{Lorentz-factor}) for the Lorentz factor is written as
\eqb
\frac{1}{\gamma^2}\left[1+\frac{\Psi_0}{\Psi}
\left(\frac{\gamma\gamma_{\rm in}^2}{\Theta ZR^2_{\rm in}}\right)^{2/3}\right]=\frac{\Psi}{\Psi_0}\frac{\beta\Theta^2}{Z^{\kappa-2}}+
\frac{\Psi_0}{\Psi}\frac 1{\Theta^2Z^2}.
 \label{gam}\eqe
The second term in the right-hand side rapidly decreases with $Z$ and becomes less than the first term at
 \eqb
Z>Z_1=\left(\beta\Theta^4\right)^{-1/(4-\kappa)}
=\left(\frac{e^{4/(\kappa-2)}}{12}\right)^{1/(4-\kappa)}\beta^{1/(\kappa-2)};
 \label{Z1}\eqe
we used Eq. (\ref{Theta}) for $\Theta$ in the last equality. Since Eq. (\ref{gam}) is valid only at $Z>Z_0$ and the ratio  \eqb
 \frac{Z_1}{Z_0}=12^{-1/(4-\kappa)}\left[\frac 12e^{2/(4-\kappa)(\kappa-2)}\right]^{2/(\kappa-2)}
 \eqe
is less than unity at $2<\kappa<3$, one can neglect the second term in the right-hand side of this equation.
Moreover one can check that $\gamma$ should grow slower than $R\propto Z$ in this case so that one can eventually neglect the second term in the square brackets. Then one finds that the Lorentz factor grows as
 \eqb
\gamma=\frac{1}{\Theta}\sqrt{\frac{\Psi_0}{\beta\Psi}}Z^{(\kappa-2)/2};
 \label{gamma_kappa>2}\eqe
which coincides with the corresponding scaling for cold flows (see Eq. (112) in ref.\cite{lyubarsky09}). As we have already
discussed at the end  of sect. 4.3, this implies that the thermal energy  is transformed into the Poynting flux, not into the kinetic energy.

One should stress that the scaling (\ref{gamma_kappa>2}) is obtained for the conical part of the jet where the curvature of the flow lines, $d^2R/dZ^2$, is
determined by small deviations from the straight line. Therefore Eq.(\ref{gamma_kappa>2})
is valid only if $R(Z)$ could be found from the governing equation (\ref{Y(Z)})
with the necessary accuracy. We show in the Appendix that this is indeed the case.

All the above scalings were obtained at the condition that the plasma is relativistically hot, $h\gg 1$. However, the expanding flow continuously cools down so that eventually
this condition is violated. In the region $Z<Z_1$, the Lorentz factor increases as $\gamma=\gamma_{\rm in}R/R_{\rm in}$ therefore Eq. (\ref{enthalpy2}) implies that the flow remains hot till $Z\sim Z_1$
provided
 \eqb
h_{\rm in}>\Theta Z_1/R_{\rm in}=%\frac{a^{1/3}}{R^{2/3}_{\rm in}\gamma^{1/3}_{\rm in}(\beta\Theta^{\kappa})^{1/(4-\kappa)}}=
R_{\rm in}^{-1}\left(\beta\Theta^{\kappa}\right)^{-1/(4-\kappa)}.
 \label{cond_h1}\eqe
Let us assume that this condition is fulfilled; then the transition to the cold flow occurs in the region $Z>Z_1$ where the Lorentz factor grows according to Eq. (\ref{gamma_kappa>2}). It follows from Eqs. (\ref{cool-radius}), (\ref{Z0}) and (\ref{gamma_kappa>2}) that the transition to the cold flow occurs at the distance
 \eqb
Z_c=\left(\frac{\sqrt{\beta}R^2_{\rm in}\gamma_{\rm in}
h^3_{\rm in}}{\Theta}\right)^{2/(\kappa+2)}.
 \eqe
The flow Lorentz factor at the transition point is
 \eqb
 \gamma_c=\left(\frac{3}{\beta^2\Theta^{2\kappa}}\right)^{1/(\kappa+2)}
 \left(R^2_{\rm in}\gamma_{\rm in}h^3_{\rm in}\right)^{(\kappa-2)/(\kappa+2)}.
 \eqe
With account of the condition (\ref{cond_h1}), one gets
 \eqb
\frac{\gamma_c}{\gamma_{\rm in}h_{\rm in}}<%a^{-2(4-\kappa)/(\kappa+2)}
\left(\frac{R_{\rm in}}{\gamma_{\rm in}}\right)^{4/(2+\kappa)}.
 \eqe
Taking into account that $R_{\rm in}<\gamma_{\rm in}$ in relativistically hot flows, one sees that $\gamma_c<\gamma_{\rm in}h_{\rm in}$, which means that most of the thermal energy is transferred not to the kinetic energy but to the Poynting flux.

When the plasma becomes cool, the flow still remains Poynting dominated. It has been shown in ref. \cite{lyubarsky09} that at $\kappa>2$, the cold flow could be accelerated according to the scaling (\ref{gamma_kappa>2}) only till a terminal Lorentz factor $\gamma_t\sim (\gamma_{\rm max}/\Theta^2)$ after which the acceleration becomes extremely slow, $\gamma\sim\gamma_t\left(\ln
Z/Z_t\right)^{1/3}$.
% \eqb
%\gamma=\left(\frac{2\sqrt{3}\gamma_{max}}{\Theta^2}\ln
%\frac{Z}{Z_t}\right)^{1/3}.
% \eqe
Therefore efficient conversion of the Poynting flux into the flow kinetic energy
could occur only if $\gamma_{\rm max}\lesssim\gamma_t$, which is equivalent to the condition
 \eqb
\gamma_{\rm max}\Theta\lesssim 1.
\label{cond_sigma}\eqe
Now we see that this statement could be generalized to hot flows. Since the cold flow could not be accelerated efficiently to $\gamma$ larger than $\gamma_t$, the final Lorentz factor
is the smallest of $\gamma_c$, $\gamma_t$  and $\gamma_{\rm max}$. If the flow was initially Poynting dominated, $h_{\rm in}\gamma_{\rm in}\ll\gamma_{\rm max}$, it becomes cool still being Poynting dominated, $\gamma_c\ll\gamma_{\rm max}$, and therefore the efficient conversion of the Poynting flux into the kinetic energy could anyway occur only if the condition (\ref{cond_sigma}) is fulfilled.

\section{Conclusions}

In this paper, we studied hot, strongly magnetized, relativistic jets at large distances from the source.
The acceleration zone of relativistic, Poynting dominated jets spans a very large range of scales therefore the processes far away from the source are of special interest. Multi-scale systems generally pose a strong challenge to numerical simulations.
On the other hand, they are suitable for asymptotic analysis. We derived equations governing the flow in the far zone; they do not contain terms that nearly cancel each other, as the general MHD equations do, and therefore they could be solved relatively easily. We concentrated on relativistically hot flows because cold, Poynting dominated jets has already been extensively studied. Our results directly generalize the results of ref. \cite{lyubarsky09} for cold flows.

Making use of the obtained asymptotic equations, we studied in detail the structure of the jet from a rigidly rotating source, $\Omega=\it const$.
The collimation and acceleration of the flow are intimately connected and determined by the distribution of the confining pressure. We adopted a power law distribution, $p_{\rm ext}\propto z^{-\kappa}$, so that the solution depends on two parameters, the index $\kappa$ showing how fast the pressure decreases, and the parameter $\beta$ defined by Eq. (\ref{beta_lin}) and showing how strong is the external pressure extrapolated to the light cylinder. We have found that hot, Poynting dominated jets are collimated exactly as in the cold case. Namely,
at $\kappa\leq 2$ the flow shape is described by a power law function ( Eq. (\ref{Ykappa<2})
for $\kappa<2$ and Eqs. (\ref{Ykappa=2a}) and (\ref{Ykappa=2b}) for $\kappa=2$) so that the jet opening angle continuously decreases.  At $\kappa>2$ the flow becomes asymptotically radial, the final collimation angle being presented by Eq. (\ref{Theta}).

When the flow expands, the fluid cools down. We have shown that at $\kappa<2$ or $\kappa=2$, $\beta>1/4$, the thermal energy is converted into the kinetic energy of the flow. On the contrary, at $\kappa=2$, $\beta<1/4$ or $\kappa>2$, the thermal energy is converted into the Poynting flux so that even the relativistically hot flow is accelerated as if it is cold. Note that acceleration regimes of cold flows are also different in these parameter ranges. At $\kappa\le 2$, the cold flows are accelerated until the equipartition between the Poynting and the plasma kinetic energy fluxes is eventually achieved. At $\kappa>2$, the flow acceleration ceases when a limiting Lorentz factor is achieved so that the flow could remain Poynting dominated \cite{lyubarsky09}. One now sees that this conclusion remains valid also for relativistically hot but Poynting dominated flows.

%after this, the flow is accelerated further extremely slowly so that the true matter dominated stage is achieved only at logarithmically large distances \cite{lyubarsky10}.

In gamma-ray bursts, the relativistic jet is formed during the collapse of star's core. In this case, the outflow is initially relativisticallly hot. Observations of the burst afterglows suggest that the final opening angle of the jet is a few degrees whereas the final Lorentz factor is at least a few hundreds. This implies $\Theta\gamma\gg 1$, which is characteristic for the case $\kappa>2$. We have shown that in this case the thermal energy is converted into the Poynting flux, not into the kinetic energy. Therefore the thermal acceleration could not help to transform the Ploynting flux into the plasma energy. Some sort of magnetic dissipation is necessary in order to utilize the electro-magnetic energy completely.
\vskip 1 cm
%\begin{acknowledgments}
This work was supported by the US-Israeli Binational Science Foundation under grant number 2006170 and by the Israeli Science Foundation under grant number 737/07.
%\end{acknowledgments}

%\appendix
\section*{Appendix. Corrections of the order of $h\gamma/\mu$ to the flux surface shape.}

We found the jet structure neglecting
$h\gamma$ as compared with $\mu$  in the Bernoulli equation
(\ref{Bernoulli2}). Then the shape of the flux surfaces is
given by Eq. (\ref{x(psi)}) where $R(Z)$ satisfies the
governing equation (\ref{Y(Z)}). The accuracy of the procedure, to within a factor $h\gamma/\mu$, is generally sufficient in the Poynting dominated jets. However, a special attention should be given to the case
$\kappa>2$ when the flow becomes radial at large distances (see sect. 4.4). The Lorentz factor of the flow
depends, according to Eq. (\ref{Lorentz-factor}), on the curvature of the flow line, $d^2R/dZ^2$. When the flow is
close to radial, the curvature is determined by small corrections to the function (\ref{cone}).
In sect. 2.4, we found the Lorentz factor of the flow in the form (\ref{gamma_kappa>2}) assuming that
the curvature of the flow lines could be found from the solution to the governing equation. Inasmuch as
this equation is valid only to within the factor $h\gamma/\mu$, we have to check whether the curvature, and consequently the Lorentz factor, thus found is correct.  %We can check applicability of the expression (\ref{gamma_kappa>2}) for the Lorentz factor of the flow by
To do so we substitute the Lorentz factor from Eq. (\ref{Lorentz-factor}) into the Bernoulli
equation (\ref{Bernoulli2}) and find the corresponding
corrections to the shape of the flux surfaces. The expression
(\ref{gamma_kappa>2}) is valid while the curvature due to this
corrections remains small as compared with the curvature (\ref{curv_vs_r})
obtained from the solution of the governing equation.

Let us present the shape of the flux surfaces as (cf. Eq. (\ref{x(psi)}))
 \eqb
X(Z,\Psi)=R(Z)\sqrt{\frac{\Psi}{\Psi_0}}\,[1+\delta(Z,\Psi)];
 \label{X_corr}\eqe
where $\delta(\Psi,Z)\ll 1$ describes corrections to the shape of the flux
surfaces due to a non-zero $h\gamma/\mu$. Substituting this expression into Eq.
(\ref{Bernoulli2}) and linearizing with respect to small $\delta$ and
$h\gamma/\mu$, one gets, with the aid of Eq. (\ref{energy1}),
% \eqb\frac{\partial\delta}{\partial\Psi}=\frac{\Omega^2h\gamma}{\eta\mu^2}. \eqe
%Making use of Eq. (\ref{energy1}), one
%writes this equation in the dimensionless form
 \eqb
\frac{\partial\delta}{\partial S}=\frac{h\gamma}{2\gamma_{\rm
max}S^2};
  \label{corr_Bern}\eqe
where
 \eqb
S=\Psi/\Psi_0;\qquad\gamma_{\rm max}=\mu(\Psi_0)=2\Omega^2\Psi_0/\eta.
 \eqe
With $\gamma$ from Eq. (\ref{gamma_kappa>2}) and $h$ from Eq. (\ref{enthalpy2}), one finds
\eqb
\frac{\partial\delta}{\partial S}=\frac{h_{\rm in}\gamma_{\rm in}^{1/3}R_{\rm in}^{2/3}}{2\Theta^{4/3}\beta^{1/3}\gamma_{\rm max}S^{7/3}Z^{(4-\kappa)/3}};
  \eqe
which yields
 \eqb
\delta=-\frac{3h_{\rm in}\gamma_{\rm in}^{1/3}R_{\rm in}^{2/3}}{8\Theta^{4/3}\beta^{1/3}\gamma_{\rm max}S^{4/3}Z^{(4-\kappa)/3}}.
 \eqe
Substituting this expression into Eq. (\ref{X_corr}) and differentiating twice
with respect to $Z$, one finds the curvature of the flux surface as
 \eqb
\frac{d^2X}{dZ^2}=\sqrt{S}\left(\frac{d^2R}{dZ^2}-
\frac{(4-\kappa)(7-\kappa)h_{\rm in}\gamma_{\rm in}^{1/3}R_{\rm in}^{2/3}}{24\Theta^{1/3}\beta^{1/3}\gamma_{\rm max}S^{4/3}Z^{(7-\kappa)/3}}\right).
 \label{curv}\eqe
Here we take into account that $R\approx\Theta Z$.

The Lorentz factor of the flow could be determined from the solution to the
governing equation only if the second term in brackets in Eq. (\ref{curv}) is small
as compared with the first one. The governing equation yields  Eq.
(\ref{curv_vs_r})  for $d^2R/dZ^2$ therefore the ratio of the first to the second terms is written as
 \eqb
\frac{24(\beta\Theta)^{4/3}\gamma_{\rm max}Z^{\frac 23(5-2\kappa)}}
{(4-\kappa)(7-\kappa)h_{\rm in}\gamma_{\rm in}^{1/3}R_{\rm in}^{2/3}}.
\label{ratio}\eqe
At the point $Z_0$ beyond which the acceleration law (\ref{gamma_kappa>2})
is valid, this ratio could be estimated, with the aid of Eqs. (\ref{Theta}) and (\ref{Z0}), as
 \eqb
\frac{2^{10/3}3^{4/3}}{(4-\kappa)(7-\kappa)}\left[\frac 1e\left(\frac 2{\kappa-2}\right)^{5-2\kappa}\right]^{4/3(\kappa-2)}\frac{\gamma_{\rm max}}{h_{\rm in}\gamma_{\rm in}^{1/3}R_{\rm in}^{2/3}}.
 \eqe
Taking into account that $h_{\rm in}\gamma_{\rm in}\ll\gamma_{\rm max}$ and $\gamma_{\rm in}\gg R_{\rm in}$ one sees that this quantity is large. If $\kappa\le 2.5$, the ratio (\ref{ratio}) remains large at $Z>Z_1$ therefore in this case the governing equation correctly describes the curvature of the filed line, which justifies the use of Eq. (\ref{gamma_kappa>2}) for the Lorentz factor of the flow. At $\kappa>2.5$, the above ratio decreases with the distance therefore eventually
Eq. (\ref{gamma_kappa>2}) could become invalid; we do not consider this case because
we are interested in strongly collimated jets whereas according to Eq. (\ref{Theta}),
the flow is only weakly collimated at $\kappa>2.5$.

An important point is that the above conclusions were obtained at the condition that the flow is relativistically hot. It is shown in ref. \cite{lyubarsky09} that in cold flows, the corrections of the order of $\gamma/\mu$
to the flux surface shape eventually become significant in $\kappa>2$ flows; then the acceleration stops.

%Note that the widely used parameter
%$\sigma$, defined as the ratio of the Poynting to the matter
%energy flux, is presented via the basic quantities as
% \eqb
%\sigma=\frac{\mu-\gamma}{\gamma}.
% \eqe

\hyphenation{Post-Script Sprin-ger}

%\bibliographystyle{myst}
%\nocite{*}

%\bibliography{mhd2}
\end{document}